\documentclass[11pt,twoside]{article}

\usepackage{asp2004}
\usepackage{epsf}
\usepackage{psfig}
\usepackage{lscape}

\begin{document}
\title{\boldmath $\sigma$ Ori E: The Archetypal Rigidly Rotating Magnetosphere}   
\author{Rich Townsend}   
\affil{Bartol Research Institute, University of Delaware, Newark, Delaware 19716, USA}    

\begin{abstract} 
I review the basic concepts of the Rigidly Rotating Magnetosphere
model for the circumstellar plasma distribution around the
helium-strong star $\sigma$ Ori E. I demonstrate that the model can
furnish a good fit to the photometric, spectroscopic and magnetic
variability exhibited by this star, and argue that the variability of
other helium-strong stars may be amenable to a similar interpretation.
\end{abstract}



\section{Introduction} \label{sec:intro}

Even among the active OB stars considered at this meeting, the B2p
star $\sigma$~Ori~E (HD~37479) is a remarkable object. The discovery
by \citet{Ber1956} of elevated helium abundances in the star's
spectrum established a new class of `helium-strong' chemically
peculiar stars, whose members today number over a dozen \citep[see,
e.g.,][and references therein]{ShoBro1990}. Likewise, the discovery by
\citet{LanBor1978} of the star's strong magnetic field furnished the
first evidence that --- quite surprisingly on account of the
\emph{absence} of envelope convection --- early-type stars can harbor
significant magnetic fields.

Yet, even more significant than this pair of `firsts' is the diverse
range of diagnostics showing periodic variability in $\sigma$ Ori
E. The star exhibits 1.19\,d modulations in its H${\alpha}$ emission
\citep{Wal1974}, its helium-line strengths \citep{PedTho1977}, its
visible and IR photometry \citep{Hes1976}, its UV diagnostics
\citep{SmiGro2001}, its 6\,cm radio emission \citep{LeoUma1993}, its
linear polarization signature \citep{KemHer1977}, and its longitudinal
field strength \citep{LanBor1978}. On account of its characteristic
double-eclipse light curves (see Oksala \& Townsend, these
proceedings), the star was for a while considered an eclipsing binary
system \citep[e.g.,][]{Hes1976,KemHer1977}. However, with the
accumulation of observational data, this interpretation of the
variability was abandoned in favor of an oblique dipole rotator model,
incorporating two co-rotating circumstellar plasma clouds situated at
the intersections between the star's magnetic and rotational equators.

\begin{figure}[t!]
\plotone{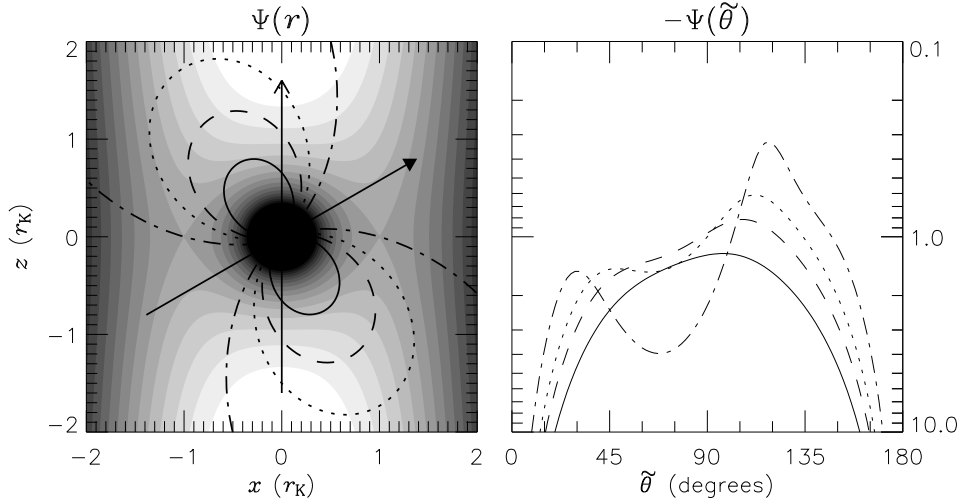}
\caption{A map of the dimensionless potential $\Psi$ in the $x$-$z$
plane (left), and the corresponding values (right) as a function of
magnetic colatitude $\tilde{\theta}$ along the four dipole field lines
plotted over the map. The arrows indicate the rotation (open
head) and magnetic (solid head) axes.} \label{fig:pot}
\end{figure}

This latter empirical picture is strongly supported by the
observations \citep[e.g.,][]{GroHun1982,Bol1987}, but it is only
recently that a matching theoretical framework has emerged. Building
on a number of previous studies \citep[see, e.g.,][]{Nak1985,
ShoBro1990, Bol1994}, \citet{TowOwo2005} have developed a new
\emph{Rigidly Rotating Magnetosphere} (RRM) model for the distribution
of circumstellar plasma around stars characterized by strong magnetic
fields and rapid rotation. In this contribution, I review the basic
concepts of the RRM model (Sec.~\ref{sec:rrm}), and demonstrate how it
is able to reproduce the photometric, magnetic and spectroscopic
variability exhibited by $\sigma$ Ori E (Sec.~\ref{sec:obs}). I also
discuss other helium-strong stars that appear amenable to a similar
interpretation (Sec.~\ref{sec:other}), examining whether $\sigma$ Ori
E can be regarded not as a one-off curiosity, but as the archetype of
a distinct class of objects whose variability arises due to an RRM.

\section{The Rigidly Rotating Magnetosphere Model} \label{sec:rrm}

In a star harboring both a magnetic field and an outflowing wind, the
interaction between field and wind plasma may be characterized
\citep{udDOwo2002} by the ratio between magnetic and kinetic energy
densities,
\begin{equation} \label{eqn:eta}
\eta \equiv \frac{B^{2}}{4\pi \rho v^{2}}.
\end{equation}
When $\eta \ll 1$, the field is completely dominated by the wind, and
is typically drawn out into a purely radial configuration. Conversely,
in the $\eta \gg 1$ limit the wind is completely dominated by the
field, to such an extent that field lines may be approximated as rigid
pipes that constrain the plasma to flow along trajectories fixed
\emph{a priori} by the magnetic topology.

In this latter limit, wind streams traveling from opposite footpoints
of a closed magnetic loop collide near the loop summit, forming a
strong shock that cools rapidly via X-ray emission
\citep[see][]{BabMon1997a,BabMon1997b}. The fate of the cooled
post-shock plasma depends on whether it is in stable or unstable
equilibrium with respect to the effective potential
\begin{equation}
\Phi(\mathbf{r}) = -\frac{G M_{\ast}}{r} - \frac{1}{2} \Omega^{2} r^{2}
\sin^{2}\theta
\end{equation}
comprising the gravitational and centrifugal potentials. If a plasma
parcel is at a local minimum of $\Phi$, as sampled along the magnetic
field line threading the parcel, then it finds itself in stable
equilibrium. In such cases, it can remain suspended above the stellar
surface for an indefinite amount of time, supported against the inward
pull of gravity by both magnetic tension and the centrifugal force
arising from enforced co-rotation.

To illustrate where such potential minima can arise,
Fig.~\ref{fig:pot} maps the dimensionless potential $\Psi \equiv
R_{\rm K} \Phi/G M_{\ast}$ in the $x$-$z$ plane. Here, $R_{\rm K}
\equiv (G M_{\ast}/\Omega^{2})^{1/3}$ is the Kepler co-rotation
radius, defined analogously to the Earth's geostationary orbital
radius. Plotted over the potential map are four selected field lines
for a dipole magnetic field whose axis lies in the same $x$-$z$ plane,
but is tilted by an obliquity $\beta = 60^{\circ}$ with respect to the
rotation axis. The dimensionless potential along these field lines, as
a function of the magnetic colatitude $\tilde{\theta}$, is plotted
beside the map.

Evidently, the potential along the outer two field lines exhibits
minima at $\tilde{\theta} \sim 70^{\circ}$. As argued above, these
minima represent points where cooled post-shock plasma can rest in
stable equilibrium, constituting a magnetosphere that co-rotates
rigidly with the star. For non-zero obliquity $\beta$, the
`accumulation surface' comprising the locus of all stable-equilibrium
points resembles a warped disk \citep[see][]{TowOwo2005}, having a
central hole of radius $\sim R_{\rm K}$. The hole arises because, as
is the case with the two inner field lines shown in
Fig.~\ref{fig:pot}, the centrifugal force close to the star is not
sufficiently strong to support material against gravity.

\begin{figure}[t!]
\plotone{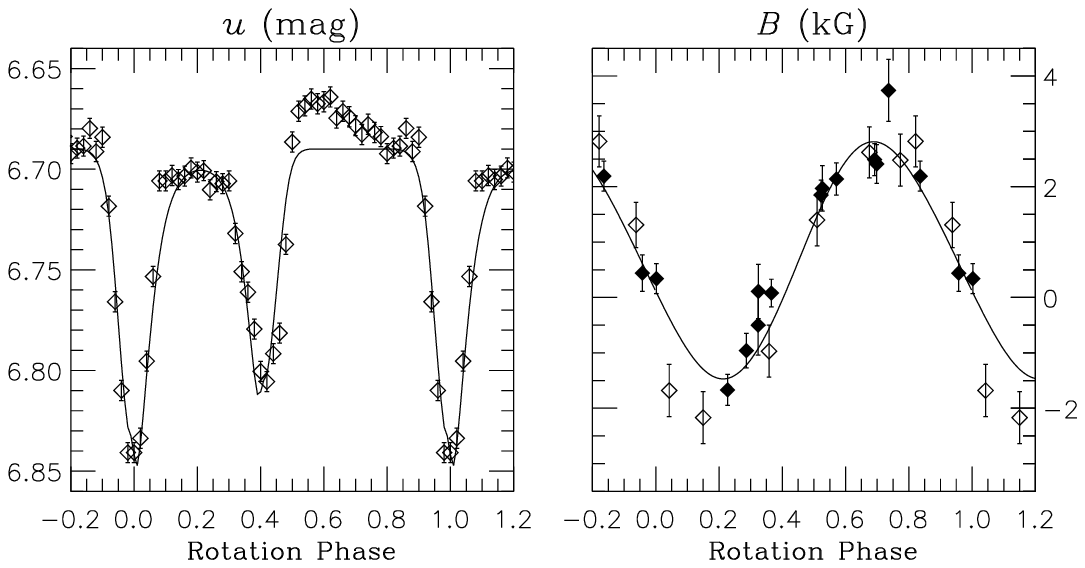}
\caption{Phased variations in the Str\"{o}mgren $u$-band flux (left)
and longitudinal magnetic field strength (right) of $\sigma$ Ori
E. Observational data are taken from \citet[][left, open
diamonds]{Hes1977}, \citet[][right, open diamonds]{LanBor1978} and
\citet[][right, filled diamonds]{Boh1987}; in both panels, the solid
line shows the relevant prediction of the RRM model.}
\label{fig:phot-field}
\end{figure}

The plasma distribution along each field line threading an
accumulation surface is dictated by the condition of
magnetohydrostatic equilibrium, and is well approximated by a Gaussian
density profile having a typical HWHM scale height $\sim
0.1\,R_{\ast}$. With steady feeding by the wind, this density profile
scales linearly with time, until some event (e.g., centrifugal
breakout; see ud-Doula, these proceedings) acts to reset the
magnetosphere to an empty state. Hence, the \emph{relative} surface
density distribution across an accumulation surface is fixed by the
mass flux into each disk element. Via the requirement of mass
conservation, this flux itself is equated to the mass-loss rate
through the corresponding magnetically-linked photospheric element,
for which simple empirical expressions exist \citep{OwoudD2004}.

Applying this approach to an oblique-dipole configuration, a key
result is that the circumstellar density distribution is sharply
peaked into two clouds, situated at the intersections between magnetic
and rotational equators. This distribution coincides closely with the
empirical picture of $\sigma$ Ori E (see Sec.~\ref{sec:intro}),
suggesting that this star is a strong candidate for application of the
RRM model. Certainly, the rigid field approximation that the model
rests upon is extremely good in the case of $\sigma$ Ori E: the star's
$\sim 10kG$ surface field and $\sim 10^{-10} M_{\odot} {\rm yr}^{-1}$
mass-loss rate yields a magnetic energy ratio (eqn.~\ref{eqn:eta})
that remains above $10^{3}$ out as far as $10\,R_{\ast}$. In the
following section, I present results confirming that the RRM model
furnishes a very good fit to the star's periodic variability.

\section{Application to \boldmath $\sigma$ Ori E} \label{sec:obs}

\citet{Tow2005} present results from a preliminary attempt at
reproducing the variability of $\sigma$ Ori E using the RRM
model. Adopting the parameters $\Omega = 0.5 \Omega_{\rm crit}$,
$\beta = 55^{\circ}$ and $i = 75^{\circ}$, and assuming a dipole field
whose center is displaced somewhat from the stellar origin, they
calculate the circumstellar density distribution. From this
distribution they then synthesize time-resolved photometric fluxes and
H${\alpha}$ emission profiles.

\begin{figure}[t!]
\plotone{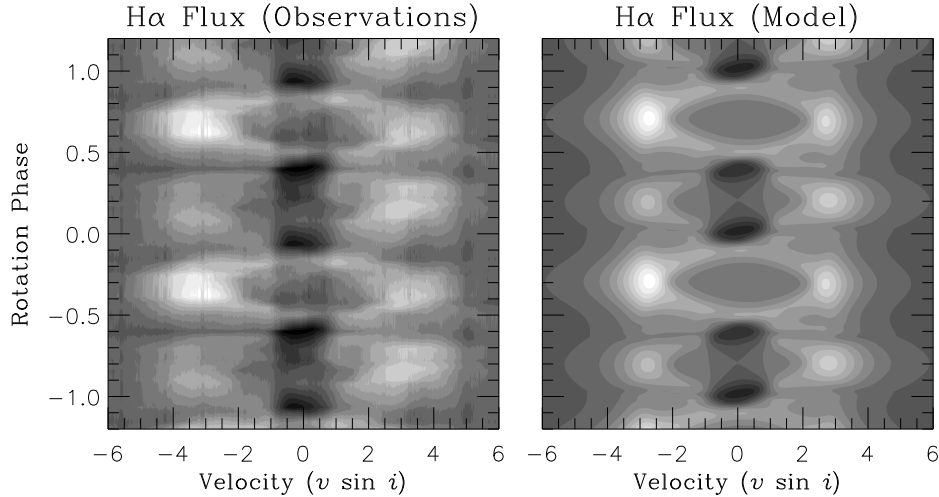}
\caption{Phased variations in the H${\alpha}$ emission profile of
$\sigma$ Ori E. White indicates a 22\% excess above the continuum, and
black a 12\% deficit. The observational data are based on 25 echelle
spectra obtained during commissioning of the \textsc{feros} instrument
\citep{Kau1999,Rei2000}; from each spectrum, a phase-dependent
synthetic absorption profile is subtracted to remove the photospheric
contribution \citep[see][]{Tow2005}.}
\label{fig:spect}
\end{figure}

Figure~\ref{fig:phot-field} compares the predicted Str\"{o}mgren
$u$-band light curve against observations of $\sigma$ Ori E. The
agreement between model and observations is encouraging; the RRM model
correctly reproduces the timing, duration and relative strength of the
twin eclipses caused when the magnetospheric clouds transit the
stellar disk. One obvious issue is the inability of the RRM model to
account for the pseudo-emission feature at phase $\sim 0.55$; however,
it appears likely that this feature arises due to the inhomogeneous
photospheric abundance distribution \citep[e.g.,][]{GroHun1982},
rather than from circumstellar plasma.

Also plotted in the figure is the predicted longitudinal magnetic
field variation, compared against the extant observational data for
the star. While not quite as good as the photometric fit, the
correspondence between model and observations is quite acceptable. If
one discards the two outlier points at phases $\sim 0.05$ and $\sim
0.15$ [which come from the original observations by
\citet{LanBor1978}] then the fit improves greatly.

Turning now to the spectroscopic variations, Fig.~\ref{fig:spect}
compares the synthetic H$\alpha$ emission profiles against recent
echelle observations of the star. In this case, the agreement between
the two is striking. The RRM model faithfully reproduces all of the
qualitative features of the double S-wave variability exhibited by the
emission. In particular, the model correctly captures the half-cycle
alternation in the strength of the blue emission peaks occurring when
the two clouds are at quadrature (phases $\sim 0.25$ and $\sim
0.75$). Yet, the alternation is absent in the corresponding red peaks,
precisely as seen in the observations. This red-blue asymmetry --- and
the unequal strengths of the eclipses in the light curve, shown in
Fig.~\ref{fig:phot-field} --- is a direct consequence of the offset
dipole field assumed in the model.

\section{Other RRM stars?} \label{sec:other}

In spite of its relative simplicity, the RRM model has proven very
successful in reproducing the variability exhibited by $\sigma$ Ori
E. What other stars might be amenable to this model? Many of the
helium-strong class should be considered as potential candidates,
since their salient parameters --- field strength, rotation rate and
mass-loss rate --- are quite similar to those of $\sigma$ Ori E.

One particular object warranting further investigation is $\delta$ Ori
C (HD~36485). \citet{Boh1987} measured a constant $\sim -3.4\,{\rm
kG}$ longitudinal field for this star, and a small projected rotation
velocity $v \sin i = 32\,{\rm km\,s^{-1}}$. Subsequent spectroscopic
observations by \citet{Boh1991} discovered twin-peaked H$\alpha$
emission that varies on a period $\sim 1.5\,{\rm d}$. While proper
modelling of the star is required, it seems not unreasonable to
hypothesize that $\delta$ Ori C is an example of a low-inclination ($i
\sim 10^{\circ}$) RRM star. The magnetic obliquity is likely
intermediate ($\sim 45^{\circ}$), since values too close to zero do
not permit the segregation of magnetospheric plasma into two clouds,
while values near $90^{\circ}$ cannot produce significant mean
longitudinal field strengths.

Another promising pole-on RRM candidate is HD~96446. \citet{Boh1987}
found a nearly-constant $\sim -1.65\,{\rm kG}$ field and an upper
limit $v \sin i \leq 30\,{\rm km\,s^{-1}}$ on the projected rotation
velocity. Photometric monitoring by \citet{MatBoh1991} revealed that
the star varies by $\sim 0.02\,{\rm mag}$ over a period $0.82\,{\rm
d}$, presumably due to rotational modulation of surface
inhomogeneities. Hence, the star is very close to being pole
on. Unfortunately, there are no H$\alpha$ profiles in the published
literature, so at this stage it is not clear whether the star actually
harbors a magnetosphere.

A final object of interest is HD~37776. \citet{ThoLan1985} reported
that the star's magnetic field is quadrapolar, on the basis of
double-wave variability in the longitudinal field strength. Recent
observations \citep[e.g.,][]{GlaGer2001} suggest an even-more complex
field configuration. One of the fortes of the RRM model is that it is
applicable to completely arbitrary magnetic topologies; in this
respect, HD~37776 represents a particularly interesting target. Once
again, however, there are no H$\alpha$ profiles in the literature,
despite the fact that the star has been studied both photometrically
and spectroscopically by a number of authors (see Mikul\'{a}\v{s}ek,
these proceedings).

Based on these three promising RRM candidates, my prejudice is to
regard $\sigma$ Ori E as the archetype of a new class of early-type
magnetic stars that possess Rigidly Rotating Magnetospheres. The task
now is to gather together the necessary data --- both archival and
from new observing campaigns --- to test how well the RRM model might
apply to these other stars.

\acknowledgements 

This research has been partially supported by US NSF grant AST-0097983
and NASA grant LTSA04-0000-0060.



\section*{Discussion}

\noindent \textbf{Henrichs}: Which data did you use to obtain the
system parameters? What diagnostics made you decide for an
off-centered dipole?

\noindent \textbf{Townsend}: The parameters are based on the
\citet{Hes1977} photometry and \textsc{feros} spectra. They shouldn't
be regarded as the `best fit' at this stage, more as a `plausible
fit'. We introduced the dipole offset to explain the differing
strengths of the primary and secondary eclipses in the light curves,
and the red-blue asymmetry in the H$\alpha$ variations.

\noindent \textbf{Chesneau}: The dipole offset, although necessary to
explain the discrepancy between the blue and red H$\alpha$ emission,
should be based only on magnetic field data, i.e. the Zeeman
signature.

\noindent \textbf{Townsend}: In an ideal world, yes --- but
unfortunately the magnetic data for $\sigma$ Ori E are rather old and
noisy, and are not sufficient either to confirm or to reject an dipole
offset.

\noindent \textbf{Owocki}: Note that the off-center dipole axis was
also invoked by Groote et al. to explain the modulations in
helium-line strengths.

\noindent \textbf{Meynet}: You mentioned that $\sigma$ Ori E was a
helium-strong star. Can you say more about the surface composition of
this star?

\noindent \textbf{Townsend}: The analysis by \citet{Rei2000}, based on
the \textsc{feros} spectra, indicates a pair of helium-rich caps that
extend $60^{\circ}$ from the magnetic poles.

\noindent \textbf{J. Bjorkman}: The density in the disk/clouds is
increasing linearly with time. Do you observe a corresponding increase
in the H$\alpha$ emission and what timescale would you expect for such
changes.

\noindent \textbf{Townsend}: Yes, there should be an increase ---
unless perhaps the back-pressure from the accumulated plasma chokes
off any further wind feeding. The typical timescale should be around a
decade, so it would be very nice within the next few years to augment
the 1998 \textsc{feros} observations with fresh data.

\noindent \textbf{\v{S}koda}: Does the H$\alpha$ variation image
(Fig.~\ref{fig:spect}) show the apparent emission (e.g., differential
spectra) or the real H$\alpha$ emission on the spectrum?

\noindent \textbf{Townsend}: The image shows the apparent emission,
after subtraction of phase-dependent synthetic absorption profiles.

\noindent \textbf{Kub\'{a}t}: Subtraction of the photospheric spectrum
is in fact a subtraction of radiation which is reprocessed in the
region where emission is formed. This is an inconsistent approach. How
does this inconsistency affect your results?

\noindent \textbf{Townsend}: Yes, there is an inconsistency in my
simple approach. However, I estimate that the error introduced by this
is unimportant when compared with other potential sources of error in
the RRM model --- so I'm not too bothered by the inconsistency.


\begin{thebibliography}{}
\bibitem[{{Babel} \& {Montmerle}(1997{\natexlab{a}})}]{BabMon1997a}
{Babel}, J., \&  {Montmerle}, T. 1997{\natexlab{a}}, \apjl, 485, 29
\bibitem[{{Babel} \& {Montmerle}(1997{\natexlab{b}})}]{BabMon1997b}
{Babel}, J., \&  {Montmerle}, T. 1997{\natexlab{b}}, \aap, 323, 121
\bibitem[{{Berger}(1955)}]{Ber1956}
{Berger}, J. 1955, Constr. Inst. Ap. Paris, A, 217
\bibitem[{{Bohlender} {et~al.}(1987){Bohlender}, {Landstreet}, {Brown}, \&
  {Thompson}}]{Boh1987}
{Bohlender}, D.~A., {Landstreet}, J.~D., {Brown}, D.~N., \&  {Thompson}, I.~B.
  1987, \apj, 323, 325
\bibitem[{{Bohlender} {et~al.}(1991){Bohlender}, {Walker}, \&
  {Bolton}}]{Boh1991}
{Bohlender}, D.~A., {Walker}, G.~A.~H., \&  {Bolton}, C.~T. 1991, \jrasc, 85,
  202
\bibitem[{{Bolton}(1994)}]{Bol1994}
{Bolton}, C.~T. 1994, \apss, 221, 95
\bibitem[{{Bolton} {et~al.}(1987){Bolton}, {Fullerton}, {Bohlender},
  {Landstreet}, \& {Gies}}]{Bol1987}
{Bolton}, C.~T., {Fullerton}, A.~W., {Bohlender}, D., {Landstreet}, J.~D., \&
  {Gies}, D.~R. 1987, in Proc. IAU Colloq. 92: Physics of Be Stars,
  {Slettebak}, A., \&  {Snow}, T.~P., eds., 82
\bibitem[{{Glagolevskij} \& {Gerth}(2001)}]{GlaGer2001}
{Glagolevskij}, Y.~V., \&  {Gerth}, E. 2001, Bull.~Special Astrophys.~Obs., 51,
  84
\bibitem[{{Groote} \& {Hunger}(1982)}]{GroHun1982}
{Groote}, D., \&  {Hunger}, K. 1982, \aap, 116, 64
\bibitem[{{Hesser} {et~al.}(1977){Hesser}, {Ugarte}, \& {Moreno}}]{Hes1977}
{Hesser}, J.~E., {Ugarte}, P.~P., \&  {Moreno}, H. 1977, \apjl, 216, 31
\bibitem[{{Hesser} {et~al.}(1976){Hesser}, {Walborn}, \& {Ugarte}}]{Hes1976}
{Hesser}, J.~E., {Walborn}, N.~R., \&  {Ugarte}, P.~P. 1976, \nat, 262, 116
\bibitem[{{Kaufer} {et~al.}(1999){Kaufer}, {Stahl}, {Tubbesing}, {Norregaard},
  {Avila}, {Francois}, {Pasquini}, \& {Pizzella}}]{Kau1999}
{Kaufer}, A., {Stahl}, O., {Tubbesing}, S., {Norregaard}, P., {Avila}, G.,
  {Francois}, P., {Pasquini}, L., \&  {Pizzella}, A. 1999, The Messenger, 95, 8
\bibitem[{{Kemp} \& {Herman}(1977)}]{KemHer1977}
{Kemp}, J.~C., \&  {Herman}, L.~C. 1977, \apj, 218, 770
\bibitem[{{Landstreet} \& {Borra}(1978)}]{LanBor1978}
{Landstreet}, J.~D., \&  {Borra}, E.~F. 1978, \apjl, 224, L5
\bibitem[{{Leone} \& {Umana}(1993)}]{LeoUma1993}
{Leone}, F., \&  {Umana}, G. 1993, \aap, 268, 667
\bibitem[{{Matthews} \& {Bohlender}(1991)}]{MatBoh1991}
{Matthews}, J.~M., \&  {Bohlender}, D.~A. 1991, \aap, 243, 148
\bibitem[{{Nakajima}(1985)}]{Nak1985}
{Nakajima}, R. 1985, \apss, 116, 285
\bibitem[{{Owocki} \& {ud-Doula}(2004)}]{OwoudD2004}
{Owocki}, S.~P., \&  {ud-Doula}, A. 2004, \apj, 600, 1004
\bibitem[{{Pedersen} \& {Thomsen}(1977)}]{PedTho1977}
{Pedersen}, H., \&  {Thomsen}, B. 1977, \aaps, 30, 11
\bibitem[{{Reiners} {et~al.}(2000){Reiners}, {Stahl}, {Wolf}, {Kaufer}, \&
  {Rivinius}}]{Rei2000}
{Reiners}, A., {Stahl}, O., {Wolf}, B., {Kaufer}, A., \&  {Rivinius}, T. 2000,
  \aap, 363, 585
\bibitem[{{Shore} \& {Brown}(1990)}]{ShoBro1990}
{Shore}, S.~N., \&  {Brown}, D.~N. 1990, \apj, 365, 665
\bibitem[{{Smith} \& {Groote}(2001)}]{SmiGro2001}
{Smith}, M.~A., \&  {Groote}, D. 2001, \aap, 372, 208
\bibitem[{{Thompson} \& {Landstreet}(1985)}]{ThoLan1985}
{Thompson}, I.~B., \&  {Landstreet}, J.~D. 1985, \apjl, 289, L9
\bibitem[{{Townsend} \& {Owocki}(2005)}]{TowOwo2005}
{Townsend}, R.~H.~D., \&  {Owocki}, S.~P. 2005, \mnras, 357, 251
\bibitem[{{Townsend} {et~al.}(2005){Townsend}, {Owocki}, \& {Groote}}]{Tow2005}
{Townsend}, R.~H.~D., {Owocki}, S.~P., \&  {Groote}, D. 2005, \apjl, 630, L81
\bibitem[{{ud-Doula} \& {Owocki}(2002)}]{udDOwo2002}
{ud-Doula}, A., \&  {Owocki}, S.~P. 2002, \apj, 576, 413
\bibitem[{{Walborn}(1974)}]{Wal1974}
{Walborn}, N.~R. 1974, \apjl, 191, 95
\end{thebibliography}
\end{document}